\documentclass[letterpaper, 10 pt, conference]{ieeeconf}
\usepackage{amssymb}
\usepackage{color}
\usepackage{bbm}

\usepackage{graphicx}
\usepackage{mathtools}
\usepackage{multirow}
\usepackage{lineno,hyperref}
\usepackage{url}
\usepackage{cleveref}
\usepackage{subcaption}
\usepackage{bm}
\usepackage{algorithm}
\usepackage{algorithmicx}
\usepackage{algpseudocode}
\usepackage{caption}
\modulolinenumbers[5]

\newtheorem{remark}{Remark}[section]

\usepackage{multirow}
\usepackage{color}
\usepackage{paralist}
\usepackage[space]{cite}

\newcommand{\ie}{\emph{i.e.}}
\newcommand{\eg}{\emph{e.g.}}
\newcommand{\R}{\mathbb{R}}

\renewcommand{\P}{\mathbb{P}}

\newcommand{\lrp}[1]{\left(#1\right)}
\newcommand{\lrs}[1]{\left[#1\right]}

\renewcommand{\t}{\theta}

\renewcommand{\d}{\delta}

\renewcommand{\i}[2]{\int_{#1}^{#2}}

\newcommand{\amin}[1]{\underset{#1}{\operatorname{arg\,min}}}

\renewcommand{\S}[3]{\sum_{\substack{#1 \\ #2}}^{#3}}

\renewcommand{\l}{\label}

\title{\LARGE \bf
Minimax Two-Stage Gradient Boosting for Parameter Estimation
}

\author{Braghadeesh Lakshminarayanan and Cristian R. Rojas
\thanks{}
\thanks{This work has been supported by the Swedish Research Council under contract number 2016-06079 (NewLEADS) and by the Digital Futures project EXTREMUM. The authors are with the Division of Decision and Control Systems, KTH Royal Institute of Technology, 100 44 Stockholm, Sweden (e-mails: blak@kth.se; crro@kth.se).
      }%
}

\begin{document}

\maketitle

\begin{abstract}
Parameter estimation is an important sub-field in statistics and system identification. Various methods for parameter estimation have been proposed in the literature, among which the Two-Stage (TS) approach is particularly promising, due to its ease of implementation and reliable estimates. Among the different statistical frameworks used to derive TS estimators, the min-max framework is attractive due to its mild dependence on prior knowledge about the parameters to be estimated. However, the existing implementation of the minimax TS approach has currently limited applicability, due to its heavy computational load. In this paper, we overcome this difficulty by using a gradient boosting machine (GBM) in the second stage of TS approach. We call the resulting algorithm the Two-Stage Gradient Boosting Machine (TSGBM) estimator. Finally, we test our proposed TSGBM estimator on several numerical examples including models of dynamical systems. 
\end{abstract}

\section{Introduction}


Most physical and economical systems can be represented by mathematical models involving unknown parameters that need to be estimated. Examples include spring-mass systems, which are modelled by second-order differential equations involving mass and spring constants, and financial systems that can be approximated by the Black-Scholes formula, depending on unknown drift and volatility parameters. To understand and simulate the behaviour of such systems, practitioners estimate these unknown parameters, which constitutes one of the important problems in system identification.


To estimate the unknown parameters of given physical system, or data generating mechanism, various estimation procedures have been developed, such as Maximum Likelihood (ML)~\cite{casella2021statistical,LehmCase98, dempster1977maximum}, Instrument Variables (IV)~\cite{Soderstrom-Stoica-83}, and Prediction Error Methods (PEM)~\cite{LJUNG1976121}.
These methods may achieve statistical consistency~\cite{LehmCase98}, however can suffer from implementation drawbacks: (i) they require user specified initialization, and (ii) the final estimation problem is often a non-convex optimization program whose solution may get stuck in local optima. For this purpose, alternative estimation procedures such as Indirect Inference~\cite{Gourieroux}, the Method of Simulated Moments~\cite{Gourieroux-Monfort-97} and the Two-Stage (TS) approach~\cite{garatti2008estimation,garatti2013new}, have been proposed in the literature.     

Among the alternative methods for parameter estimation, the TS approach has several advantages: (i) no explicit likelihood computation is needed, and (ii) it can be posed as a simple convex optimization program by carefully designing the second stage, for which minimization algorithms do not get trapped into local optima. The TS approach is based on a supervised learning approach, where one simulates a large number of observations corresponding to different values of the unknown parameters, and these observations are used to train a supervised learning algorithm to predict the true values of unknown parameter. The TS approach consists of two stages: (i) a data compression stage and (ii) a function approximation stage. In the data compression stage (first stage), a large number of observations is compressed into a smaller set of samples, and in the second stage these compressed samples along with their corresponding values of the unknown parameters are fed as a training set to a supervised learning algorithm. 

The theoretical justification of the TS approach to parameter estimation was first developed in~\cite{BLCRCDC2022}. Towards this end, its authors provided two statistical frameworks for which TS can be derived. The two frameworks considered are: (i) \emph{Bayes} and (ii) \emph{Minimax} estimation. The derivations of these frameworks were done for the case where the data generating mechanism creates independent and identically distributed (i.i.d.) samples, even though they can be easily extended to parameter estimation of dynamical systems where the data is non-i.i.d. Under the i.i.d. assumption, the first stage can be constructed from quantiles of the observations. It has been demonstrated via numerical example in~\cite{BLCRCDC2022} that the TS approach gives reliable Bayes and minimax estimates of the unknown parameters.

In this paper, we focus on the minimax TS framework. This framework is practically and theoretically attractive because, unlike Bayes estimators, minimax estimators are ``robust" to prior knowledge. However, the minimax TS approach derived in \cite{BLCRCDC2022} suffers from high computational time required to obtain reliable estimates, due to the fact that (i) an optimization algorithm should be run for each sample of unknown parameters, and the number of such samples is large, and (ii) most of the constraints in the epigraph formulation, used in~\cite{BLCRCDC2022} to solve the minimax problem, are inactive, which make the solver run very slow. In this paper, we save computation time by deploying a Gradient Boosted Machine (GBM) in the second stage as the supervised learning algorithm, which is a non-linear function approximator that is constructed based on decision or regression trees~\cite{schapire2003boosting,friedman2001greedy,friedman2002stochastic}, and by carefully designing the cost function of GBM we solve the minimax problem more efficiently. 

In particular, our contributions are:
 \begin{itemize}
    \item use a Gradient Boosted Machine (GBM) as the supervised learning algorithm in the second stage of TS approach;
    \item modify the evaluation metric of GBM using a soft-max approximation so that the min-max objective can be efficiently solved;
    \item numerically validate the minimax statistical framework on several examples, including dynamical systems where the data generation mechanism is non-i.i.d.
\end{itemize}

The rest of the paper is organized as follows:

In Section~\ref{sec: setup} we define the problem setup. Section~\ref{sec: prelims} outlines the preliminaries. In Section~\ref{sec: Minimax_TSGBM}, we propose the Two-Stage Gradient Boosting Machine (TSGBM) method for estimating unknown parameters, and evaluate TSGBM on numerical examples in Section~\ref{sec: Simulation}. Finally, we conclude the paper in Section~\ref{sec: conclusion}. 

\section{Problem Setup} \label{sec: setup}

\begin{figure}[!h]
    \centering
    \includegraphics[width=0.5\linewidth]{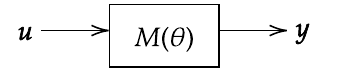}
    \caption{A data generating mechanism $M(\theta)$. $\boldsymbol{u}$ is an input to $M(\theta)$ and $\boldsymbol{y}$ is its corresponding output.}
    \label{fig: Cartoon_1}
\end{figure}

Consider a data generating mechanism as described in Figure~\ref{fig: Cartoon_1}. Let $M(\theta)$ be a model, or \emph{data generating mechanism}, that generates an observation $\boldsymbol{y}$ upon receiving an input $\boldsymbol{u}$, where $\boldsymbol{u} \in \mathcal{U} \subseteq \mathbb{R}^{p}$ is a $p$-dimensional real vector, and $\boldsymbol{y} \in \mathcal{Y} \subseteq \mathbb{R}^{r}$ is an $r$-dimensional real vector. Here, $\mathcal{U}$ and $\mathcal{Y}$ are called \emph{input space} and \emph{observation space} respectively. $\theta \in \Theta \subseteq \mathbb{R}^d$ is an unknown parameter vector in a parameter space $\Theta$ that has an explicit influence on the distribution of the output generated by $M(\theta)$.

The goal is to estimate the unknown parameter $\theta$ given a set of inputs $\{\boldsymbol{u}_i\}_{i=1}^N$ and their corresponding outputs $\{\boldsymbol{y}_i\}_{i=1}^N$. We assume that an observation $\boldsymbol{y} \in \mathcal{Y}$ has an underlying probability distribution $\mathbb{P}(\boldsymbol{y} \vert \theta, \boldsymbol{u})$, which is parameterized by $\theta$ and a known input $\boldsymbol{u}$. Under this assumption, the goal of estimating the unknown parameter translates to designing an \emph{estimator} or \emph{decision rule} $\delta\colon \mathcal{U} \times\mathcal{Y} \to \Theta$ of the true parameter $\theta$ such that the \emph{risk} 
\begin{align} \label{eq: risk}
R(\theta,\delta) &= \mathbb{E}_{\boldsymbol{y} \sim \mathbb{P}(\cdot \vert \theta, \boldsymbol{u})} [L(\theta, \delta(\boldsymbol{y})]
\end{align}
is minimized, where $L\colon \Theta \times \Theta \to \mathbb{R}_{0}^{+}$\footnote{$\mathbb{R}_{0}^{+}$ is the set of all non-negative real numbers} is a \emph{loss} function that evaluates the cost of estimating the unknown parameter as $\delta(\boldsymbol{y})$ when its true value is $\theta$. Due to the dependence of the risk on $\theta$, which is a priori unknown, the estimation problem can be approached in at least two different ways~\cite{casella2021statistical, berger2013statistical}, which lead, respectively, to Bayes and minimax rules.

In this paper, we focus on minimax rules, where a decision rule $\delta_{\text{minimax}}^{*}$ minimizes $\max\limits_{\theta \in \Theta} R(\theta, \delta)$. That is,
\begin{equation}\label{eq: minimax_rule}
    \delta_{\text{minimax}}^{*} = \amin{\delta \in \Delta}  \max\limits_{\theta \in \Theta} R(\theta, \delta),
\end{equation}

\section{Preliminaries} \label{sec: prelims}

In this section we briefly revise the main concepts behind the TS approach, its minimax variant, and gradient boosting machines. 

\subsection{TS Approach} \label{subsec:ts}

The TS approach to parameter estimation is an inverse supervise learning method where several values of the unknown parameter $\theta$, namely $\{\theta_i\}_{i=1}^{M_{\theta}}$, are sampled from a probability distribution over $\Theta$, and for each sample $\theta_i$ a large number of observations $\boldsymbol{y}_i = (y_1^{(i)},\ldots,y_N^{(i)})^T$ is generated (see Fig~\ref{fig: cartoon_2}). The first stage of the TS approach consists in compressing the observations $\boldsymbol{y}_i$ into a smaller set of samples $\boldsymbol{\alpha}_i=(\alpha_1^{(i)},\ldots,\alpha_n^{(i)})^T$ where $\alpha_j^{(i)} \in \mathbb{R}$ for $j \in \{1,\ldots,n\}$, and $n \ll N$. The compression is achieved via a fixed function $h\colon \mathbb{R}^N \to \mathbb{R}^n$, that is, $h(\boldsymbol{y}_i) = \boldsymbol{\alpha}_i$. In the case of i.i.d. observations, a suitable choice for $h(\boldsymbol{y}_i)$ is given by some quantiles of $\boldsymbol{y}_i$, whereas in the case of non-i.i.d. observations, $h(\boldsymbol{y}_i)$ may correspond, \eg, to the coefficients of an estimated AR($n$) model. This compression step aims to mitigate the effects of measure concentration; see \cite{BLCRCDC2022} for details. Once these compressed samples are obtained, in a second stage a conventional supervised learning algorithm~\cite{Shalev-Shwartz-Ben-David-12} such as Kernel regression, Deep Neural Networks (DNNs) or Gradient Boosting, is used with $\{(\boldsymbol{\alpha}_i,\theta_i)\}_{i=1}^{M_{\theta}}$ as the training set. That is, a non-linear function approximation is learned using a supervised learning algorithm, denoted by $g\colon \mathbb{R}^n \to \mathbb{R}^d$, which then outputs an estimate of $\theta$.

\subsection{Minimax TS Estimator} \label{subsec: minimax_ts }
Now, we briefly review the minimax TS approach developed in~\cite{BLCRCDC2022}. The minimax estimator, denoted by $\delta_{\text{minimax}}^{*}$, is given by~\eqref{eq: minimax_rule}. Using the explicit expression for the risk $R(\t,\d)$, we obtain
\begin{equation}
    \max_{\t \in \Theta} \, R(\t,\d) = \max_{\t \in \Theta} \i{\mathcal{Y}}{} L\lrp{\t,\d(\boldsymbol{y})} \P(\boldsymbol{y}|\t) d\boldsymbol{y}.
\end{equation}

Further, the integral in the above optimization problem can be approximated by a Monte Carlo sum, by drawing several samples of observation $\boldsymbol{y}$ for a fixed $\theta$. This gives
\begin{equation} \label{eq: approx_minmax_1}
\max_{\t \in \Theta} \, R(\t,\d) \approx \max_{\t \in \Theta} \lrs{\dfrac{1}{M_\mathbf{y}} \S{j=1}{}{M_\mathbf{y}} L(\t,\d(\mathbf{y}_{j,\t}))},
\end{equation}
where $M_{\boldsymbol{y}}$ is the number of Monte Carlo samples of $\boldsymbol{y}$ generated from $\mathbb{P}(\cdot \vert \theta)$, \ie, $\{\boldsymbol{y}_{j,\t}\} \stackrel{i.i.d.}{\sim}{\P(\cdot \vert \t)}$. $\Theta$ can be potentially infinite, due to which~\eqref{eq: approx_minmax_1} may be computationally intractable. Hence, we use scenario approach~\cite{Calafiore-Campi-2006} to discretize $\Theta$ by sampling several values of $\theta$ using a proposal distribution $s$ over $\Theta$, i.e., $\theta_i \stackrel{i.i.d.}{\sim} s$. After discretization, \eqref{eq: approx_minmax_1} becomes


\begin{equation}\label{eq: approx_minmax_2}
\min_{\d \in \Delta} \, \max_{i = 1, \dots, M_\t} L_i(\d),
\end{equation}
where $L_i(\d) := \S{j=1}{}{M_\mathbf{y}}L({\t}_i,\d(\mathbf{y}_{ij}))$ and $\mathbf{y}_{ij} := \mathbf{y}_{j, \t_i}$. Using an epigraph formulation~\cite[page~134]{boyd_vandenberghe_2004}, we obtain the final approximate optimization problem as
\begin{align}\l{eq:min_max_cvx}
\begin{array}{cl}
\displaystyle \min_{\d \in \Delta, t \in \R} & t  \\
\text{s.t.} & L_i(\d) \leq t, \quad i = 1, \dots, M_{\t}.
\end{array}
\end{align}

\begin{remark}
While minimax TS was proposed in~\cite{BLCRCDC2022} for i.i.d. data generating mechanisms, this approach can also be applied when the data is non-i.i.d. Taking a closer look at~\eqref{eq: approx_minmax_1}, the Monte Carlo samples are in fact i.i.d., \ie, $\{\boldsymbol{y}_{j,\t}\} \stackrel{i.i.d.}{\sim}{\P(\cdot \vert \t)}$, although each sequence $\boldsymbol{y}_{j,\t}=(y_{j,1}^{(\t)},\ldots,y_{j,N}^{(\t)})^T$ is non-i.i.d., that is, the $k^{th}$ observation $y_{j,k}^{(\t)}$ may depend on the the past $k-1$ values $y_{j,1}^{(\t)},\ldots,y_{j,k-1}^{(\t)}$. 
\end{remark}

\subsection{Gradient Boosting} \label{subsec: gb}

Gradient boosting is a machine learning paradigm for functional estimation. In function estimation problems, we have a training sample $\{(\boldsymbol{x}_i,\boldsymbol{y}_i)\}_{i=1}^H$, and the goal is to estimate a function $F$ that maps $\boldsymbol{x}_i \in \mathcal{X}$ to $\boldsymbol{y}_i \in \mathcal{Y}$. For simplicity, we will assume in this section that $\mathcal{Y} \subseteq \mathbb{R}$.
The estimated function, denoted by $F^*$, is obtained by solving the optimization problem
\begin{equation} \label{eq: boosting_opt_problem}
F^* = \amin{F \in \mathcal{F}} \frac{1}{H} \sum_{i=1}^H L(y_i,F(\boldsymbol{x}_i)),
\end{equation}
where $L$ is the loss function defined in~\eqref{eq: risk} and $\mathcal{F}$ contains functions of the additive form
\begin{equation}\label{eq: boost_additive_form}
    F(\boldsymbol{x}) = \sum_{p=1}^P w_p k(\boldsymbol{x};\beta_p),
\end{equation}
where $k_p(\cdot;\beta_p)$ are \emph{weak models} or learners that are simple functions parameterized by $\beta_p$, \eg, $k(\boldsymbol{x};\beta_p) = \beta_p^T \boldsymbol{x}$. Substituting~\eqref{eq: boost_additive_form} in~\eqref{eq: boosting_opt_problem} yields an optimization problem in $\{w_p\}_{p=1}^P$ and $\{\beta_p\}_{p=1}^P$, which is usually solved in a forward ``stage-wise" fashion, where we start with an initial guess $F_0$ and then for $p=2,\ldots,P$,
\begin{equation}\label{eq: boost_final_opt}
    w_p,\beta_p = \amin{w,\beta} \sum\limits_{i=1}^H L(y_i,F_{p-1}(\boldsymbol{x}_i)+ w k(\boldsymbol{x}_i;\beta)),
\end{equation}
with $F_p(\boldsymbol{x}) = F_{p-1}(\boldsymbol{x}) + w_p k(\boldsymbol{x};\beta_p)$. 
%

Gradient Boosted Machine (GBM) is a tree-based gradient boosting algorithm that uses a regression tree as a weak learner $k(\boldsymbol{x},\beta)$. In particular, at each step $p$ in the stage-wise optimization~\eqref{eq: boost_final_opt}, a regression tree partitions the input $\boldsymbol{x}$ into $B$-disjoint leafs, denoted by the set $\{R_{jp}\}_{j=1}^B$, \ie, $k(\boldsymbol{x},\beta)$ is a regression tree of depth $C$ where the $i^{th}$ entry of $\boldsymbol{x}$ falls into one of these leaves. Thus, $\beta_p$ is replaced by the set $\{R_{jp}\}_{j=1}^B$. It has been shown in~\cite{friedman2002stochastic} that~\eqref{eq: boost_final_opt} can be equivalently solved by
\begin{align} \label{eq: gbdt_eq}
\gamma_{jp} &= \amin{\gamma} \sum\limits_{\boldsymbol{x}_i \in R_{jp}} L(y_i,F_{p-1}(\boldsymbol{x}_i) + \gamma),\\
F_{p}(\boldsymbol{x}) &= F_{p-1}(\boldsymbol{x}) + \eta \gamma_{jp} \mathbf{1}\{\boldsymbol{x} \in R_{jp}\}, \nonumber
\end{align}
where $\mathbf{1}\{A\}$ denotes the indicator function of the set $A$, and $0 < \eta \leq 1$ is the step size. 

There are several open source libraries such as XGBoost~\cite{xgboost}, CatBoost~\cite{dorogush2018catboost}, LightGBM~\cite{Ke2017LightGBMAH} that implement GBM.
Each library has its own hyperparameters that characterize a regression tree.

\section{Minimax TSGBM} \label{sec: Minimax_TSGBM}
In this section, we propose the new minimax Two-Stage Gradient Boosted Machine (TSGBM) algorithm for parameter estimation. Towards that end, we use LightGBM in the second stage of TS, which is a gradient boosting library developed by Microsoft~\cite{Ke2017LightGBMAH}. In particular, we are interested using it to minimize $\max_{\theta \in \Theta} R(\theta, \delta)$, where $\delta$ here is the non-linear map to be delivered by LightGBM. Recall from Section~\ref{subsec: minimax_ts } that the approximate minimax risk is given by
\begin{equation}\label{eq: minimax_cost}
    \min\limits_{\delta} \max_{i=1, \dots, M_{\theta}} L_i(\delta).
\end{equation}

Using LightGBM in the context of the minimax TS approach to parameter estimation requires minimizing the ``custom loss'' $\max_{i=1, \dots, M_{\theta}} L_i(\delta)$.
Such a loss does not have closed form expression, so it cannot be directly used for LightGBM.
Therefore, we use a soft-max approximation~\cite{jcook} for $\max_{i=1, \dots, M_{\theta}} L_i(\delta)$ which is given by
\begin{equation}\label{eq: soft_max}
    \max_{i=1, \dots, M_{\theta}} L_i(\delta) \approx \dfrac{\log\left( \sum\limits_{i=1}^{M_{\theta}} \exp(K L_i(\delta)) \right)}{K},
\end{equation}
where $K$ is chosen large enough (in the order of $10^{3}-10^{4}$).

\begin{remark}
In case the data generating mechanism produces i.i.d. or stationary samples, it is possible to set $M_{\boldsymbol{y}}=1$ in~\eqref{eq: approx_minmax_1} by choosing $N$ sufficiently large; this can be justified from the Ergodic Theorem (see, \eg, \cite[Lemma B.1, page~548]{Soderstrom-Stoica-89}). In our examples we thus choose $M_{\boldsymbol{y}}=1$ and $N \approx 10^{4}$.
\end{remark}

Algorithm~\ref{alg: TSGBM} describes the implementation of TSGBM. The inputs to this algorithm are the proposal distribution $s$ and a collection of tuning parameters of LightGBM, which includes
learning rate ($l_r$), number of iterations ($itr$), maximum depth of a regression tree ($B\_max$), number of leaves ($l\_num$), fraction of data used ($bg\_fr$), minimum number of data points in a leaf ($d\_l\_min$), and amount of $\ell_1$ regularization ($\lambda$).
\begin{table*}[h]
\centering
\scalebox{0.9}{
  \begin{tabular}{|c|c|c|c|c|c|c|c|c|c|c}

    \hline
      \multicolumn{2}{|c|}{\textbf{True Values}} & \multicolumn{2}{|c|}{\textbf{CRLB}} & \multicolumn{2}{c|} {\textbf{MSE, Minimax}} & \multicolumn{2}{c|} {\textbf{MSE, Minimax TSGBM}}\\
    \cline{1-8}
     \text{$\eta$} & \text{$\gamma$} & \text{${\eta}$} & \text{${\gamma}$} & \text{$\hat{\eta}$} &\text{$\hat{\gamma}$}& \text{$\hat{\eta}$} &\text{$\hat{\gamma}$} \\
    \hline\hline
      $2$ & $2$ & $1.11\times10^{-4}$ & $2.43\times10^{-4}$ & $2.58\times10^{-4}$ & $5.77\times10^{-2}$ & $0.62\times10^{-4}$ & $2.17\times10^{-4}$  \\ \hline
     $2$ & $8$ & $6.93\times10^{-6}$ & $3.89\times10^{-3}$ & $1.11\times10^{-5}$ &  $5.61\times10^{-2}$ & $0.36\times10^{-5}$ & $0.44\times10^{-2}$ \\ \hline
      $4$ & $2$ & $4.43\times10^{-4}$ & $2.43\times10^{-4}$ & $6.74\times10^{-4}$ &  $1.05\times10^{-1}$ & $3.76\times10^{-4}$ & $2.11\times10^{-4}$ \\ \hline
      $4$ & $8$ & $2.77\times10^{-5}$ & $3.89\times10^{-3}$ & $3.84\times10^{-5}$ &  $6.40\times10^{-2}$ & $3.86\times10^{-5}$ & $0.52\times10^{-2}$  \\ \hline
        $8$ & $2$ & $1.77\times10^{-3}$ & $2.43\times10^{-4}$ & $2.26\times10^{-3}$ &  $1.89\times10^{-1}$ & $2.37\times10^{-3}$ & $2.16\times10^{-4}$ \\ \hline
  \end{tabular}
}
\caption{MSE of the orignal minimax TS~\cite{BLCRCDC2022} and minimax TSGBM estimators of the scale and shape parameters, and their corresponding CRLBs.}
\label{table:crlb}
\end{table*}

The output of Algorithm~\ref{alg: TSGBM} is $\delta_{\text{TSGBM}} = g \circ h$, which is the desired TSGBM estimator. Here, $h$ is the data compression function and $g$ is the function approximation provided by LightGBM. To evaluate the performance of this estimator, its Mean Square Error (MSE) is computed via Algorithm~\ref{alg: TSGBM_MSE}. The inputs to  Algorithm~\ref{alg: TSGBM_MSE} are $\delta_{\text{TSGBM}}$, $\tilde{\theta}_{\text{test}}$ and MC: $\tilde{\theta}_{\text{test}}$ is a specific value of the unknown parameter at which the MSE is evaluated by running MC times Algorithm~\ref{alg: TSGBM}, each time returning an estimator $\delta_{\text{TSGBM}}$ based on a different training sample. 

\begin{remark}
In case the parameter vector $\theta$ is $d$-dimensional (with $d>1$), 
Algorithm~\ref{alg: TSGBM} is implemented for each entry of $\theta$ separately.
Likewise, Algorithm~\eqref{alg: TSGBM_MSE} is implemented for each dimension of the unknown parameter. 
\end{remark}

\begin{remark}
    To account for non-linearities in the true system, we propose a non-linear transformation (monomials of degree upto two) of the compressed data, which will then be a part of the training set to GBM. However, the dimension of the transformed data will be much less compared to $N$.  
\end{remark}

\begin{algorithm}[httb!]
  \caption{TSGBM Estimator}
\begin{algorithmic}[1]
 \State \textbf{Input}: $s$, $K$, tuning parameters
\State Generate $\theta_i \sim s(\cdot), \, i=1,\ldots,M_{\theta}$
 \For{$i=1,\ldots,M_{\theta}$}
 \State $\boldsymbol{y}_i \leftarrow M(\theta_i)$ \Comment{Model $M(\theta)$ as data simulator}
 \State $\boldsymbol{\alpha}_i \leftarrow h(\boldsymbol{y}_i)$ \Comment{Data compression step}
 \EndFor
 \State Construct non-linear features $\phi(\boldsymbol{\alpha}_i)$, where $\phi\colon \mathbb{R}^n \to \mathbb{R}^m$, where $n<m\ll N$
  \State \text{Define the soft-max loss as in}~\eqref{eq: soft_max}
 \State \text{Train LightGBM with} $\{(\phi(\boldsymbol{\alpha}_i),\theta_i)\}_{i=1}^{M_{\theta}}$ as the training set and~\eqref{eq: soft_max} as loss function
  
 \State \textbf{Output}:  $g$ \Comment{Function learned by LightGBM}
 \State \textbf{Final estimator:}  $\delta_{\text{TSGBM}} := g \circ h$.
 \end{algorithmic} 
 \label{alg: TSGBM}
\end{algorithm}

\section{Simulation Study} \label{sec: Simulation}
In this section we demonstrate the performance of TSGBM via numerical simulations that include both i.i.d. and non-i.i.d. data generating mechanisms. A detailed description of these mechanisms is provided in Sections~\ref{subsec: Weibull}, \ref{subsec: dyna_1} and \ref{subsec: dyna_2}.

\begin{algorithm}[httb!]
  \caption{MSE for TSGBM}
\begin{algorithmic}[1]
 \State \textbf{Input}: $\delta_{\text{TSGBM}}$, $\tilde{\theta}_{\text{test}}$, MC
\State $\text{Sum} \leftarrow 0$ 
\For{$m=1,\ldots,$ MC}
 \State $\tilde{\boldsymbol{y}}_{\text{test}} \leftarrow M(\tilde{\theta}_{\text{test}})$ \Comment{Model $M(\theta)$ as data simulator}
 \State $\tilde{\boldsymbol{\alpha}}_{\text{test}} \leftarrow h(\tilde{\boldsymbol{y}}_{\text{test}})$ \Comment{Data compression step}
 \State Construct non-linear features $\phi(\tilde{\boldsymbol{\alpha}}_{\text{test}})$
 \State $\widehat{\tilde{\theta}}_{\text{test}} = \delta_{\text{TSGBM}}(\tilde{\boldsymbol{y}}_{\text{test}})$
 \State $\text{Sum} \leftarrow (\widehat{\tilde{\theta}}_{\text{test}}-\tilde{\theta}_{\text{test}})^2$
 \EndFor
 \State $\text{MSE} \leftarrow \frac{\text{Sum}}{\text{MC}} $ 
 \State \textbf{Output}: $\text{MSE}$. 
  

 \end{algorithmic} 
 \label{alg: TSGBM_MSE}
\end{algorithm}

For the numerical simulations, the following setup is considered: The TSGBM estimator($\delta_{\text{TSGBM}}$) is implemented using Algorithm~\ref{alg: TSGBM}, with $M_{\theta}=M_{\text{train}}$ and $K=10^3$. For each $\theta_i$ in Algorithm~\ref{alg: TSGBM}, we generate observations $\boldsymbol{y}_i$ of dimension $N \times 1$. Unless otherwise stated, we do not construct non-linear features $\phi(\boldsymbol{\alpha}_i)$ in Step 7 of Algorithm~\ref{alg: TSGBM}, in which case training set in Step 9 of Algorithm~\ref{alg: TSGBM} will be $\{(\boldsymbol{\alpha}_i,\theta_i)\}_{i=1}^{M_{\theta}}$. For TSGBM, the function approximator $g$ is defined via a regression tree. To generate scatter plots for the estimated vs. true values of the parameter, we draw fresh samples of $\theta$ from the proposal distribution $s$, \ie, $\tilde{\theta}_i \sim s(\theta)$, and collect them in the set $\{\tilde{\theta}_i\}_{i=1}^{M_{\text{test}}}$. We compute the MSE according to Algorithm~\ref{alg: TSGBM_MSE}, where the values of $\tilde{\theta}_{\text{test}}$ used are listed in Tables~\ref{table:crlb}-\ref{table: mse_tsgbm_dyna_2_r4} as ``True Values".

\subsection{Weibull Distribution} \label{subsec: Weibull}
We first demonstrate the performance of minimax TSGBM on an i.i.d. data generating mechanism. Specifically, we consider a Weibull distribution whose probability density function is parameterized by two parameters, namely, \emph{scale} ($\eta$) and \emph{shape} ($\gamma$). It has been shown in~\cite{BLCRCDC2022} that the minimax TS approach gives reliable estimate for $\eta$. However, estimates for the shape parameter $\gamma$ when its true value is small are not reliable, in the sense that its MSE is significantly larger when compared to the asymptotic Cram\'er-Rao lower bound (CRLB).

However, using TSGBM, \ie, with LightGBM in the second stage, shows significant improvement in the MSE for $\gamma$, and also for the scale parameter $\eta$ the MSE is comparable to that outlined in~\cite{BLCRCDC2022}. This has been highlighted in Table~\ref{table:crlb}. Also, the training time of minimax TSGBM is considerably smaller compared to the minimax TS algorithm developed in~\cite{BLCRCDC2022}, which runs CVXPY~\cite{diamond2016cvxpy} for each training sample of $\theta$ (and the number of such training samples required for the original minimax TS method to be reliable was fairly large). Minimax TSGBM, on the other hand, does not suffer from this issue, since it directly uses a soft-max maximization objective as the custom loss during its training and minimizes this loss for the entire training sample, \ie, it performs batch training. 

\begin{figure*}[h]
\begin{minipage}{0.3\textwidth}
\centering
\includegraphics[width=0.75\linewidth]{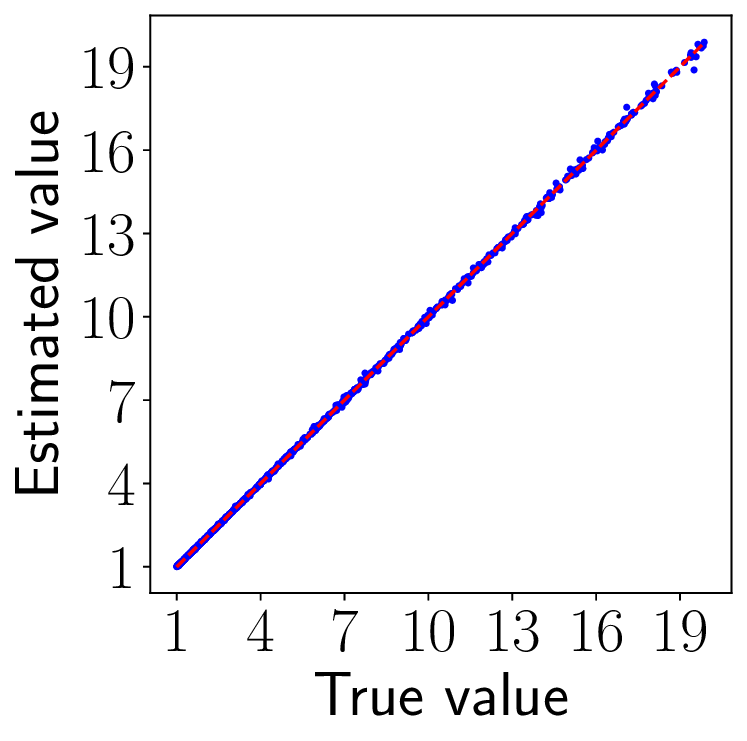}
    \caption{Scatter plot of estimated vs. true values of scale parameter ($\eta$); see Section~\ref{subsec: Weibull}.}
\label{fig: weibull_scale}
\end{minipage}
\hfill
\begin{minipage}{0.3\textwidth}
\centering
\includegraphics[width=0.9\linewidth]{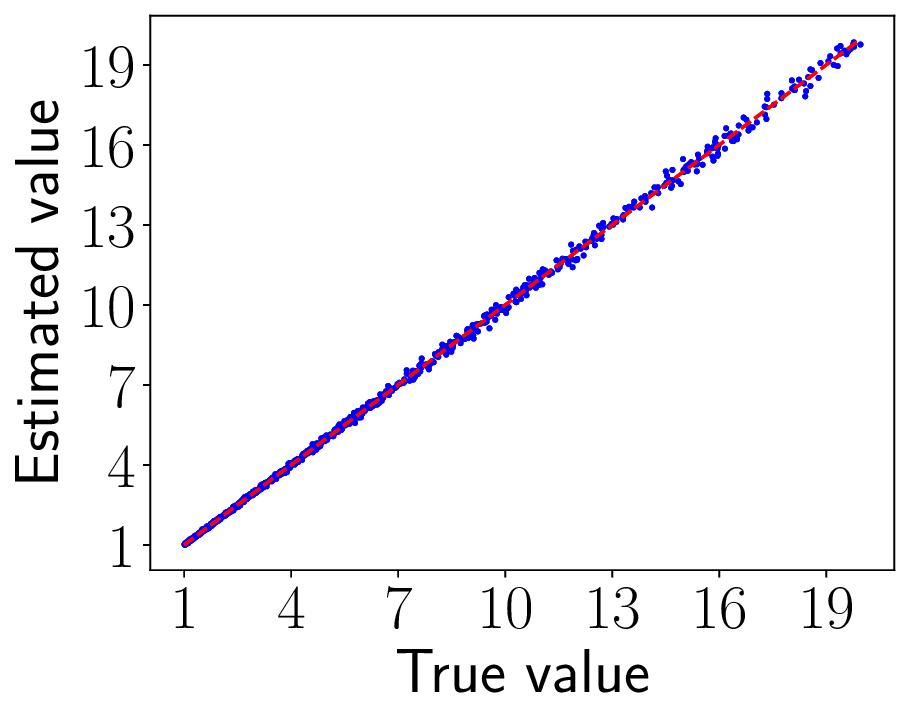}
\caption{Scatter plot of estimated vs. true values of shape parameter ($\gamma$); see Section~\ref{subsec: Weibull}.}
\label{fig: weibull_shape}
\end{minipage}
\hfill
\begin{minipage}{0.3\textwidth}
    \centering
    \includegraphics[height=0.75\linewidth,width=1\linewidth]{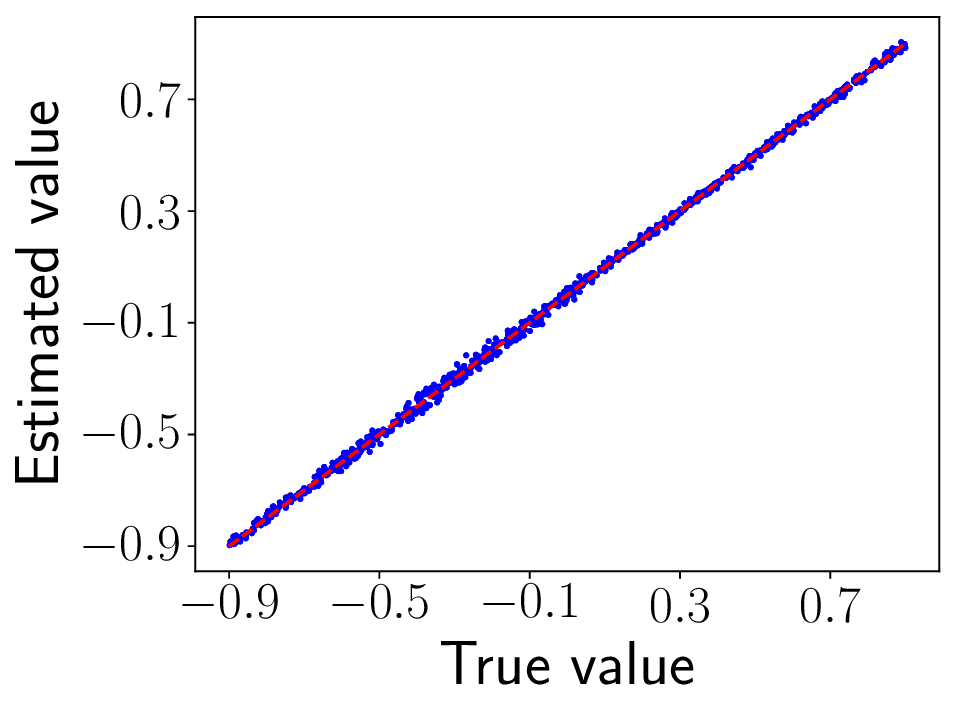}
    \caption{Scatter plot of estimated vs. true values of $a$; see Section~\ref{subsec: dyna_1}.}
    \label{fig: dyna_1_tsgbm}
\end{minipage}
\end{figure*}

For the numerical simulations, the following choices are made: (i) $h$ computes quantiles from order statistics of the observations ($n=5$), for both the original TS and TSGBM-based estimators. To compute non-linear features, $\phi$ was taken as described in~\cite{BLCRCDC2022} for $\eta$ and $\gamma$, (ii) The functional approximator $g$ was taken to be linear regression for the original TS estimator, while for TSGBM, we use LightGBM with parameters: $l_r$= $0.1$, $itr$ = $10^3$, $B\_max$= $5$, $l\_num$ = $16$,  $bg\_fr$ = $1$, $d\_l\_min$ = $20$, and  $\lambda$ = $10^{-4}$, (iii) For both TS estimators, the proposal distribution for $\eta$ and $\gamma$ is an uninformative prior~\cite{BLCRCDC2022} over $[1,20]$. Also, MC$=1000$ for MSE computations  and (iv) For the original minimax TS estimator, $M_{\text{train}} = 1000$ samples are considered, while for TSGBM, $M_{\text{train}} = 10^4$ since it is more computationally efficient and it gives reliable estimates compared to the original approach.


The scatter plots of estimated vs. true parameter shown in Figures~\ref{fig: weibull_scale}, \ref{fig: weibull_shape} complement the results given in table~\ref{table:crlb}, demonstrating that minimax TSGBM yields reliable estimates. 



\subsection{Single Parameter Dynamical System} \label{subsec: dyna_1}
We now consider a single parameter discrete-time state-space model from \cite{garatti2008estimation} that is described as
\begin{align} \label{eq: sim_dynamic_1}
    x_1(k+1) &= a x_1(k) + v_{11}(k), \nonumber \\
    x_2(k+1) &= x_1(k) + a^2 x_2(k) + v_{12}(k), \\
    y(k) &= a x_1(k) + x_2(k) + v_2(k),  \nonumber
\end{align}
where $x_1(k)$ and $x_2(k)$ are hidden states and $y(k)$ is the output at time $k$. $v_{11}(k)$ and $v_{12}(k)$ are the additive process noises and $v_2(k)$ is the observation noise at time $k$. $v_{11} \sim \mathcal{N}(0,1)$, $v_{12} \sim \mathcal{N}(0,1)$ and $v_2 \sim \mathcal{N}(0,0.01)$ are mutually uncorrelated Gaussian noises. 

\begin{table}[h!]
    \centering
  \begin{tabular}{|c|c|}
    \hline
      \multicolumn{1}{|c|}{\textbf{True Values ($a$)}} &  {\textbf{MSE}}\\
    \cline{1-2}
    $0.1$ & $1.08 \times 10^{-4}$ \\
    $0.2$ & $1.01 \times 10^{-4}$  \\
    $0.3$ & $8.12 \times 10^{-5}$  \\
    $0.4$ & $7.24 \times 10^{-5}$ \\
    $0.5$ & $1.02 \times 10^{-4}$ \\
    \hline
  \end{tabular}
\caption{MSE of TSGBM for different values of $a$; see Section~\ref{subsec: dyna_1}.}
\label{table: mse_tsgbm_dyna_1}
\end{table}
For the above state-space model \eqref{eq: sim_dynamic_1}, the parameter of interest is $a$ and we use minimax TSGBM to estimate it. For this purpose, we set $M_{\text{train}} = 10^4$, $N = 10^4$, $h$ as the coefficients of an estimated AR(5) model, $\phi$ as the collection of all monomials of degree at most $2$, $M_{\text{test}}=10^3$, $s$ as a uniform distribution over $[-1,1]$, and $\theta=a$. LightGBM is used with parameters: $l_r$ = $0.05$, $itr$ = $10^3$, $B\_max$= $4$, $l\_num$= $8$, $bg\_fr$= $0.9$, $d\_l\_min$ = $30$, and $\lambda$ = $10^{-3}$.


Figure~\ref{fig: dyna_1_tsgbm} shows a scatter plot of estimated vs. true values of $a$. We see that the estimates are quite reliable, in the sense that the relation between the true and estimated values is almost linear.


Next, we compute the MSE of TSGBM by evaluating the trained estimator for 5 different values of $a$ over $1000$ Monte Carlo runs in Table~\ref{table: mse_tsgbm_dyna_1}.


\begin{table*}
    \centering
    \footnotesize
    
    \begin{minipage}{0.3\textwidth}
  \begin{tabular}{|c|c|c|c|c|c|}
  \hline 
      \multicolumn{2}{|c|}{\textbf{True Values}}  &  \multicolumn{2}{|c|}{\textbf{MSE}}\\
      \hline
    \text{$a$} & \text{$b$} &\text{$\hat{a}$} & \text{$\hat{b}$}\\
    \cline{1-4}
    $-0.5$ & $0.45$ & $6.13 \times 10^{-4}$ & $3.83 \times 10^{-4}$\\
    $-0.3$ & $0.55$ & $2.59 \times 10^{-4}$ & $2.89 \times 10^{-4}$ \\
    $-0.1$ & $0.65$ & $1.19 \times 10^{-4}$ & $1.13 \times 10^{-4}$ \\
    $0.1$ & $0.75$ & $7.07 \times 10^{-5}$ & $4.75 \times 10^{-5}$ \\
    $0.3$ & $0.85$ & $1.89 \times 10^{-4}$ & $3.57 \times 10^{-5}$ \\
    \hline
  \end{tabular}
\caption{MSE of TSGBM for different values of $a \in [-0.5,0.5]$ and $b \in [0.45,0.9]$; see Section~\ref{subsec: dyna_2}.}
\label{table: mse_tsgbm_dyna_2_r2}
    \end{minipage}
    \hspace{9em}
    \begin{minipage}{0.3\textwidth}
  \begin{tabular}{|c|c|c|c|c|c|}
  \hline 
      \multicolumn{2}{|c|}{\textbf{True Values}}  &  \multicolumn{2}{|c|}{\textbf{MSE}}\\
      \hline
    \text{$a$} & \text{$b$} &\text{$\hat{a}$} & \text{$\hat{b}$}\\
    \cline{1-4}
    $0.1$ & $0.5$ & $1.74 \times 10^{-4}$ & $4.05 \times 10^{-4}$ \\
    $0.2$ & $0.6$ & $1.22 \times 10^{-4}$ & $1.6 \times 10^{-4}$\\
    $0.3$ & $0.7$ & $2.12 \times 10^{-4}$ & $8.68 \times 10^{-5}$\\
    $0.4$ & $0.8$ & $2.21 \times 10^{-4}$ & $4.78 \times 10^{-5}$\\
    $0.5$ & $0.9$ & $1.27 \times 10^{-3}$ & $1.42 \times 10^{-5}$\\
    \hline
  \end{tabular}
\caption{MSE of TSGBM for different values of $a \in [0.1,0.5]$ and $b \in [0.1,0.9]$; see Section~\ref{subsec: dyna_2}.}
\label{table: mse_tsgbm_dyna_2_r4}
    \end{minipage}
\end{table*}

\begin{figure*}[h!]
\begin{minipage}{0.2\textwidth}
\includegraphics[width=1\linewidth]{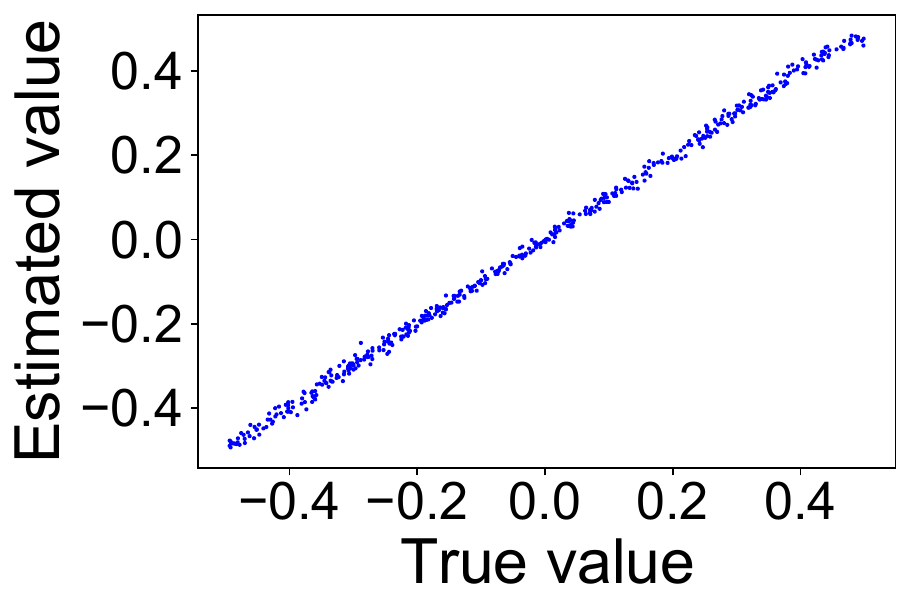}
\caption{Scatter plot of estimated vs. true values for $a \in [-0.5,0.5]$; see Section~\ref{subsec: dyna_2}.}
\label{fig: scatter_dyna_2_r4_a}
\end{minipage}
\hfill
\begin{minipage}{0.2\textwidth}
\includegraphics[width=1\linewidth]{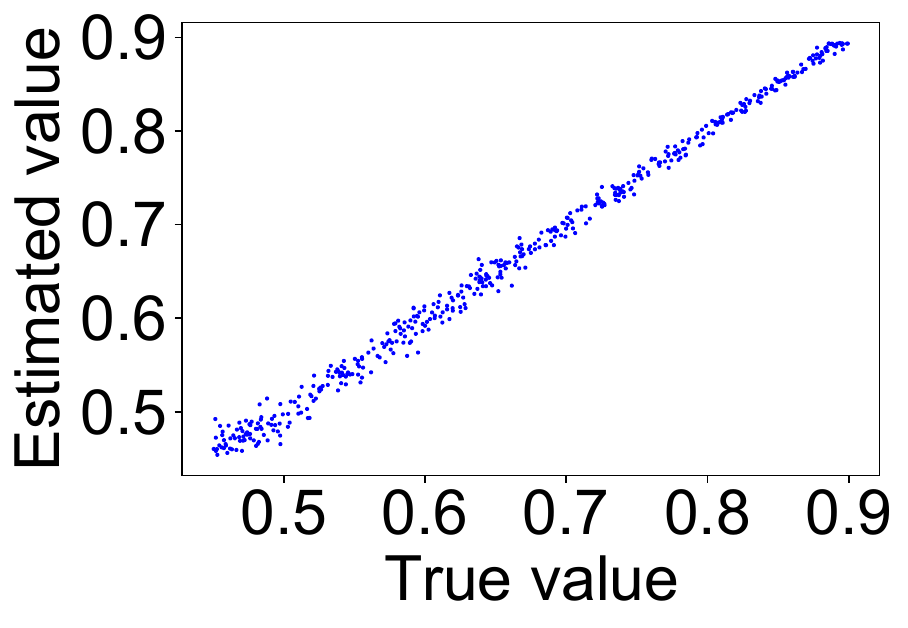}
\caption{Scatter plot of estimated vs. true values for $b \in [0.45,0.9]$
; see Section~\ref{subsec: dyna_2}.}
\label{fig: scatter_dyna_2_r4_b}
\end{minipage}
\hfill
\begin{minipage}{0.2\textwidth}
\centering 
\includegraphics[width=1\linewidth]{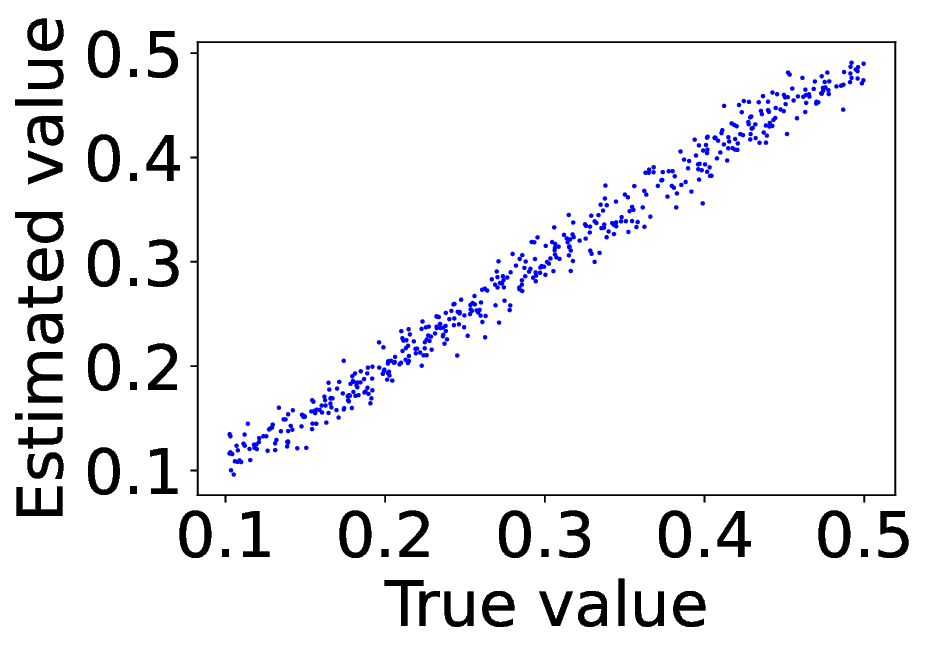}
\caption{Scatter plot of estimated vs. true values for $a \in [0.1,0.5]$; see Section~\ref{subsec: dyna_2}.}
\label{fig: scatter_dyna_2_r2_a}
\end{minipage}
\hfill
\begin{minipage}{0.2\textwidth}
\centering 
\includegraphics[width=1\linewidth]{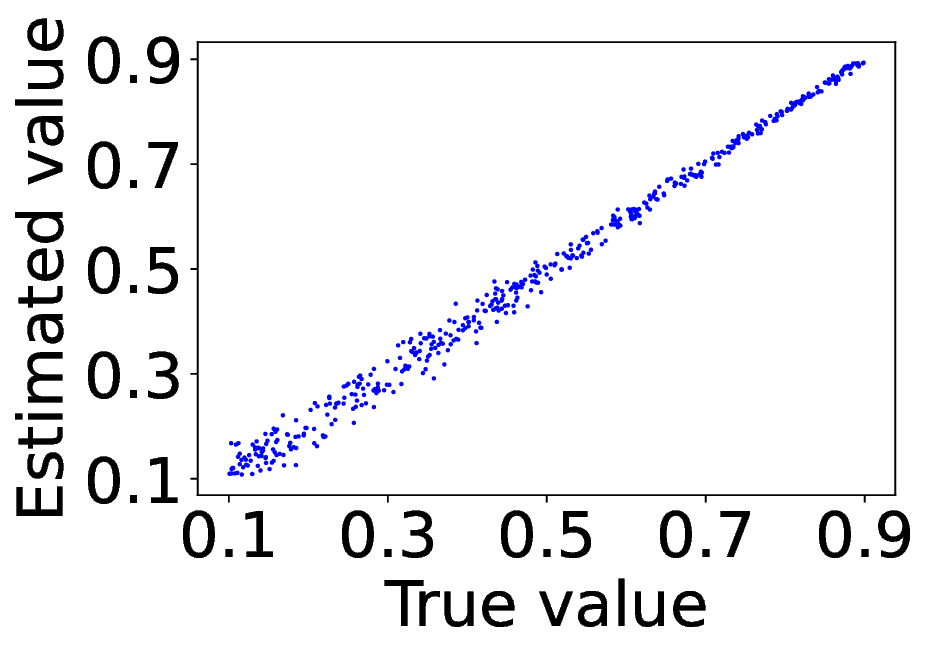}
\caption{Scatter plot of estimated vs. true values for $b \in [0.1,0.9]$; see Section~\ref{subsec: dyna_2}.}
\label{fig: scatter_dyna_2_r2_b}
\end{minipage}
\end{figure*}



Based on the results of this section, we can conclude that the minimax framework of the TS approach is indeed applicable for dynamical systems, where the data generating process is not i.i.d, and TSGBM gives very reliable estimates.

\subsection{Stochastic Volatility Model} \label{subsec: dyna_2}
In this subsection, we consider a discrete-time stochastic volatility model from~\cite{COURTS2023110687} that is described by the equations
\begin{align} \label{eq: sim_dynamic_2}
    x_{k+1} &= a + b x_{k} + v_k, \nonumber \\
    y_{k} &= \sqrt{e^{x_k}} r_k,
\end{align}
where $v_k \stackrel{i.i.d.}{\sim} \mathcal{N}(0,1)$, $ r_k \stackrel{i.i.d.}{\sim} \mathcal{N}(0,1)$ and $v_k, r_k$ are mutually uncorrelated. In this model, $a,b$ are the unknown parameters that we need to estimate. Let $\theta = (a \ b)^T$ be the parameter vector of model~\eqref{eq: sim_dynamic_2}. As one may notice, model~\eqref{eq: sim_dynamic_2} has multiplicative noise in its observation. Hence, using an estimated AR($n$) process as the compression stage of the TS approach is not a good idea, since an AR($n$) model assumes that noise is added to the output. In order to obtain an additive noise structured for \eqref{eq: sim_dynamic_2}, we transform the observations by first squaring $y_k$ and then applying the logarithm. This leads to the transformed observations
\begin{equation*}
    \bar{y}_k= x_k + \bar{r}_k,
\end{equation*}
where $\bar{r}_k = \log(r_k^2)$.

Now, $\bar{y}_k$ has an additive noise structure, and hence an AR($n$) model can be used in the first stage of TS. Therefore, the final state-space model equivalent to~\eqref{eq: sim_dynamic_2} is
\begin{align} \label{eq: sim_dynamic_2_eq}
    x_{k+1} &= a + b x_{k} + v_k \nonumber \\
    \bar{y}_k &= x_k + \bar{r}_k.
\end{align}
%


We will use this model, instead of \eqref{eq: sim_dynamic_2}, to estimate $\theta = (a \ b)^T$ using the TS approach.


For estimating $a$ and $b$, we consider ranges of $a$ and $b$ to be (i) $a \in [-0.5,0.5]$ and $b \in [0.45,0.9]$ and (ii) $a \in [0.1,0.5]$ and $b \in [0.1,0.9]$.
Figures~\ref{fig: scatter_dyna_2_r4_a}-\ref{fig: scatter_dyna_2_r2_b} show scatter plots of estimated vs. true values of the unknown parameters $a$ and $b$ of the stochastic volatility model. Tables~\ref{table: mse_tsgbm_dyna_2_r2}, \ref{table: mse_tsgbm_dyna_2_r4} list MSE of TSGBM for both $a$ and $b$. For these figures and tables, we have set $M_{\text{train}}= 10^4$, $N= 3 \times 10^4$, $M_{\text{test}} = 10^3$, $s$ as a uniform distribution over the ranges of $a,b$ considered (see below), and $h$ as the coefficients of an estimated AR(5) model. Also, we have considered a non-linear feature map $\phi$ consisting of all monomials of degree at most $2$, and the following parameters for LightGBM: $l_r$ = $0.05$, $itr$ = $10^4$, $B\_max$ = $6$, $l\_num$ = $8$, $bg\_fr$ = $0.9$, $d\_l\_min$= $30$, and $\lambda$ = $10^{-4}$. From Figures~\ref{fig: scatter_dyna_2_r4_a}-\ref{fig: scatter_dyna_2_r2_b} and Tables~\ref{table: mse_tsgbm_dyna_2_r2}, \ref{table: mse_tsgbm_dyna_2_r4}, we see that TSGBM provides fairly reliable estimates of $a$ and $b$ with minimal training effort, as it might not be the case with DNNs or recurrent neural networks.

\section{Conclusion} \label{sec: conclusion}
In this paper we have developed a computationally efficient minimax implementation of the TS approach, and we have numerically demonstrated that it is applicable for dynamical systems with non-i.i.d. observations. In particular, the new algorithm, called TSGBM, uses gradient boosting as the supervised learning algorithm in the second stage of TS, and it provides reliable estimates of unknown parameters with minimal training effort, in contrast to using, say, deep neural networks.

It should be remarked that choosing the first stage of TS to appropriately capture the temporal and probabilistic structure of the observations is very important, but its careful design, in the non-i.i.d. case, is left for future work.  
\bibliography{References}



\end{document}